\documentclass[12pt,oneside,english]{article}

\usepackage{epsf}
\usepackage[T1]{fontenc}
\usepackage[latin1]{inputenc}
\usepackage{babel}
\usepackage{setspace}
\makeatletter

\begin{document}

\section*{\bf    Internet websites statistics  expressed in the framework of 
the Ursell-Mayer   cluster formalism  } 
\makeatother

\bigskip \bigskip 

\centerline{\large  D.  Bar$^{a}$} 
{\bf $^a$Department of Physics, Bar Ilan University, Ramat Gan,
Israel }

\bigskip \bigskip \bigskip 

   \noindent 
   {\it  We show that it is possible to  generalize the Ursell-Mayer 
cluster formalism  
so that it may cover also the  statistics of Internet websites. 
Our starting point is the introduction of an extra variable that is assumed 
to take account, as will be explained,  of the nature of the Internet 
 statistics. We then show, following the arguments in Mayer, that one may obtain 
a phase transition-like phenomena.}

 \bigskip \bigskip 
  
\noindent
{\bf \underline{Keywords}}: Internet, Ursell-Mayer formalism, Phase transition, Combinatorics

\bigskip \bigskip \bigskip \noindent 

\protect \section{\bf INTRODUCTION \label{sec1}}

\bigskip 

The use of   the Internet as a necessary tool for  easing the application  of an 
increasing number of diverse tasks, and not only for web surfing,  is fastly  
growing.   
There has been  lately an ongoing research    that discusses the Internet topology 
\cite{Faloutsos,Albertt} where  use is  made of the 
fractal \cite{Mendelbrot,Havlin1} and the percolation theories 
\cite{Havlin2,Aharony,Newman}.  
  In  these works  the Internet is regarded as random network 
\cite{Havlin1,Havlin2} and the websites as its building blocks. \par 
We focus  our attention here on the unique nature of the websites  
 which enables the  possible existence in them  
of any number of links that refer to other sites that  may be downloaded 
by a single click of the computer mouse.      
  As known,  any  fractal is built by repeated 
iteration (see Aharony in \cite{Havlin2}) of some unique natural "microscopic" 
growth rule which is rather special in   the 
 Internet case.  This is because  not only the forms of the constituent 
websites,  the identity and connectivity \cite{Havlin1,Havlin2,Aharony} 
of their links  
depend  only on the programmers that  write the relevant softwares   
 but also the growth of the web itself depends exclusively upon them.  
   Thus, for  taking  into account the nature   of  programming  that enables 
one to display  on the computer  screen practically {\it anything}  
    we introduce a special variable, denoted in the following by $s$, 
    that corresponds  
to  the spatial variable $r$ which is used to discuss the statistics of the $N$ 
particle system. The character of this variable $s$ will be discussed in the following 
section.  \par 
 We note that similar situations arise in the  discussion of some mathematical 
\cite{Gelfand} and physical \cite{Lax} 
situations for which one have to add a special variable that takes account of some 
nonphysical properties of the discussed systems.   
  For example, in the functional generalization \cite{Horwitz} of Quantum 
Mechanics  the  generalized  Hilbert space 
(the Lax-Phillips one \cite{Lax}) is obtained by adding an extra variable to the conventional 
Hilbert space. 
A similar  addition of  an extra variable  
   has been proposed  
 by Parisi-Wu-Namiki \cite{Parisi1,Namiki1} in their stochastic Quantization theory 
  which 
assumes that some stochatic process \cite{Kannan} occurs in the  
extra dimension.  \par    
As known, any web site contains one or more web pages and each of these may 
include one or more links to other places on the same page or to other sites 
or  other web servers. The user that clicks, through the keyboard,  on any 
link (the highlighted addresses (URL)) downloads its relevant site to the screen. The number of the links in 
a web site may be small (or even zero for the unlinked web sites) or it may 
be very large.   We do not consider here the secondary  
links that may be found on the sites referred to by the first links. \par 
We discuss the connectivity \cite{Faloutsos,Albertt,Havlin1,Havlin2} 
among the Internet websites where by this term 
we mean  the amount and degree of the interconnection among the sites that compose 
the Internet network. Thus, a large connection among the sites of 
 the Internet denotes a corresponding large connectivity and vice versa.
Our aim in this work is to show that if we consider large clusters of 
mutually linked sites 
then adding even a very small amount of connecting links results in 
a disproportionally  
large  growth  of the total connectivity \cite{Havlin1,Havlin2} 
of these clusters. 
We use, for that purpose, a generalized version of  Ursell \cite{Ursell} 
and Mayer 
\cite{Mayer,Reichl} cluster formalism.  This generalization is obtained 
 by the introduction  of 
the remarked  extra variable that takes account of   
the unique Internet statistics. We note that similar discussion in Ursell
\cite{Ursell} 
 and Mayer \cite{Mayer} with regard to the large clusters  of particles results in 
demonstrating phase-transition phenomena \cite{Ursell,Mayer,Reichl}.    
\par 
 In Section 2 we discuss the Internet using  
the partition function method  \cite{Mayer,Reichl,Huang}  
where the relevant  "configuration
integral"  is discussed by generalizing  
the cluster formalism of Ursell \cite{Ursell} and Mayer \cite{Mayer}. This
generalization is imposed by the unique nature of the Internet websites which
allows, as will be shown, more than one kind of linkage between the sites. 
Thus,  the
typical numbering procedure and the combinatorics of the Ursell-Mayer method,
which was originally formulated for discussing the $N$-particle system, has to
be appropriately expanded. In Section 3 we discuss and generalize the relevant
statistical integrals and in Section 4 we show the mentioned phase transition of
the connectivity among the Internet websites. Note that our 
discussion in this work is general in that we do not specify the nature and identity of 
the relevant sites.      
 \bigskip     
\bigskip  
\bigskip \noindent 
\pagestyle{myheadings}
\markright{ THE ADAPTATION OF THE  URSELL-MAYER CLUSTER FORMALISM ....}

\bigskip \bigskip  

 \protect \section{THE ADAPTATION OF THE  URSELL-MAYER CLUSTER FORMALISM  FOR 
 THE INTERNET WEBSITES \label{sec2}}

 \protect \subsection{The definitions of the "position",  the "distance" and 
 the "potential" between
 websites} 
 
 \bigskip \bigskip

 We  show in this work that the connectivity
 \cite{Havlin1,Havlin2} in large $N$-site clusters, which is determined by the
 links in all the sites, is so sensitive to these connecting links  that
 adding even a small number of them to the cluster results in a giant increase
 of its overall connectivity. This large growth of the connectivity, 
 compared to the small addition of links that causes it,  constitutes a phase
 transition which should be discussed in the appropriate terms and terminology. 
 For that purpose 
we adapt    a generalized version of the virial expansion of the
equation of state \cite{Reichl,Huang} which is discussed by using the cluster
formalism developed  by Ursell \cite{Ursell} and 
Mayer \cite{Mayer}. We assume that the web sites system discussed here is 
 composed of $N$ sites where $N$ is generally a large number. We note that 
 when one discuss the potential energy of  the $N$ 
particle system in the
configuration space \cite{Reichl,Huang} the relevant variable is the distance 
$r_{ij}$ 
between any two particles $i$ and $j$ which depends only on their positions.  
Thus,  the potential between them 
$u(r_{ij})$ is assumed to be effective  for small ranges of $r_{ij}$ and  
 vanishes when $r_{ij}$ grows. \par  Here, also  for the $N$ site system 
  we denote by $u(s_{ij})$ the "potential" between the two sites $i$ and $j$ which
  depends upon the  "distance", denoted here by $s_{ij}$,  between them.  
   This ``distance''  signifies  how much these sites differ from each other 
 so that it is shorter the more ``similar''  they are and
 longer for unrelated sites. The distance between $i$ and $j$ 
may be assigned a Quantitative
 aspect  by taking into account the number of common links to both so that 
the  larger this number is the  shorter  this distance becomes. 
  That is, each site $i$ is characterized by all its
 links so that its "position" in $s$ space may be written as
 $s_i(s_{i1},s_{i2},s_{i3},\ldots)$, where, $s_{i1}$, $s_{i2}$, $s_{i3}$ etc
 denote the links in $i$ to   the sites 1, 2, 3, etc.  Thus, as the
real  positions  in configuration space  are measured relative to the 
origin (zero values for coordinates) 
so the "positions"  of the sites here  are "measured"  
 relative to the unlinked site which has zero link. That is, the more linked  
 some  site is the larger is its "distance" from the "origin"  
 (the unlinked site). 
  In this context we may use Figure 1 in order to make  
 this point clearer. Thus, the unlinked sites denoted [7] and [8] at the right hand side of
 the planar diagram of Figure 1 are at the largest "distance" from each other
 and from the other sites of the figure. On the other hand, the doubly linked
 sites [15], [16], [23] and [24] in this figure have a very small "distance" 
 
 from each other.     
  In such a manner one may write the "distance" between
  $i$ and $j$, for example,   as  $s_{ij}(s_{i1},s_{i2},\ldots, 
 s_{j1},s_{j2},\ldots)=s_i(s_{i1},s_{i2}\ldots)-s_j(s_{j1},s_{j2},\ldots)$.  
 Thus, we  may define a 
 "potential" $u(s_{ij})$,  in analogy to the $N$ particle system,
 for example, as 
 \begin{equation}  u(s_{ij})=\frac{1}
 {s_i(s_{i1},s_{i2},\ldots)-s_j(s_{j1},s_{j2},\ldots)} \label{e1} \end{equation} 
The potential $u(s_{ij})$ between the sites $i$ and $j$  does not
have a physical meaning but only a statistical one. This is because it
depends  upon the distance $s_{ij}$ between these sites which is, as remarked,
determined  not only by their mutual links (to each other) 
but also by all the other shared and common 
 links which may be very large in number. Thus, this potential have a
statistical meaning (see the discussion after Eq (\ref{e6}) and (\ref{e7}) 
and the Appendix) 
which is expressed using Combinatorical analysis. \par
The potential $u(s_{ij})$ from the last equation, as for the corresponding $u(r_{ij})$,  
is effective for  small ``distances'' $s_{ij}$ and vanishes when $s_{ij}$ grows.  
However,  in contrast to the $N$ particle system in
 which the distance $r_{ij}$ between any two members   $i$ and $j$  is sharply and
 uniquely determined  by their  positions  only here  the "distance" 
  between any two  sites is also  
 characterized, as has been remarked, by other links to other sites that are
 common to both, besides the links that refer to each other.    Moreover, since,
 as remarked, the linking among the sites are effected through the highlighted
 places (upon which one press with the mouse pointer to download the linked site
 to the screen) a site
 may have such a link to another one  whereas the second have no link to 
 the first.   
 Thus, in order to
 discuss appropriately the potential    between 
 any two sites $i$ and $j$ one must differentiate between  four different 
 situations;  
 (1): $i$ has a link to $j$ and $j$ has no one to $i$.  (2): 
 $j$ has a link to $i$ and $i$ has no one to $j$.  (3): both $i$ and $j$ have no
 links to each other.  (4): both $i$ and $j$  have links to each other in which
 case the connectivity between them is larger than in cases 1 and 2. 
 \protect \subsection{The "Thermodynamical" discussion and representation of the
 Internet}
 
 \bigskip \bigskip
 
 The mentioned possible different kinds of linkings between the sites influences the
 standard expressions and formulas of the Ursell-Mayer expansion \cite{Mayer} in
 such a manner that  the application of it to the Internet statistics becomes,
 as will be shown,  delicate and complicated.  First, we note that the use of
 physical thermodynamical methods for discussing the Internet structure enables
 one not only to use the variable $s$ in an analogous manner to the spatial
 variable $r$ but also to use other thermodynamical quantities. We note in this
 context that the use of Thermodynamics terminology and terms for discussing
 other (nonphysical) branches of science may be found   in the general
 literature (see, for example,
 \cite{Bennett,Feynman}). Thus, we may  
 discuss
  an appropriately defined partition function $Z(s)$ as well as 
 the related "pressure" $P(s)$  and the "free energy" $F(s)$  in the linked cluster of
 websites. We show by applying these fundamental concepts to the Internet that
 the  remarked phase
 transition of the cluster's connectivity  may be  related to  the internetic  
 "pressure" $P(s)$ of its sites.   This will be demonstrated for large clusters 
 of 
 sites  which are doubly connected to each other as in (4) above (see the
 discussion after Eq (\ref{e1}))  in which
 case adding even a small number of links results in a disproportional enormous
 strengthening of the overall connectivity. We first define  the appropriate
 "partition function" $Z(s)$ that determines 
      the states of the $N$ site 
ensemble and the correlation among them  
\begin{equation} 
 \label{e2} Z(s)=\sum_{N=0}^{N = \infty}f(s)C_N(V(s))  \end{equation} 
 As seen, we use the same expression for $Z(s)$ 
as used in  
 \cite{Reichl,Huang} except for the dependence upon the variable $s$.  
  The $f(s)$ is generally an 
exponential function that does not
 depend upon the potential $V(s)$ and 
   $C_N(V(s))$ is the ``configuration integral'' in $s$ space which is 
 given here by the same form, except for the $s$ dependence, as that of the
 particle system \cite{Reichl,Mayer} 
 \begin{equation} \label{e3} C_N(V(s))=\int \int \ldots \int e^{-\beta_sV(s)}ds_1
 ds_2 \ldots  ds_i \ldots ds_N,       
 \end{equation}   The variable $s$  denotes that we discuss  Internet 
 sites  and  each of the differentials $ds_1ds_2 \ldots ds_N$ 
signifies  an infinitesimal "volume" element in $s$-space. Thus, if we assume, 
analogously to 
the configuration space, that this space may be projected into 
three  axes denoted by $a1$, $a2$ and $a3$ then the $i$ differential in Eq 
(\ref{e3}) may be written as $ds_i=ds_{i_{a1}}ds_{i_{a2}}ds_{i_{a3}}$.  Now,  
since  $s_{ij}$  denotes, as  
remarked, the 
``distance'' between the sites $i$ and $j$ (see discussion before Eq (\ref{e1}))
 then we may regard the integrals in Eq (\ref{e3}) as ranged over all 
"distances"  from a reference site which corresponds  to the origin of 
the configuration 
space. This  
 reference site is, as remarked  before Eq (\ref{e1}),   
 the unlinked site  (with zero links).   The $\beta_s$   in the
 exponent has dimension of "inverse energy" and corresponds to the
 $\beta$ of the 
  particle ensemble \cite{Mayer,Reichl,Huang}.   Thus, from Eqs 
 (\ref{e2})-({\ref{e3}), and analogously to the $N$ particle system
 \cite{Mayer}, 
 one may define the appropriate "pressure"  $P(s)$ and "free energy" $F(s)$ as
    \begin{equation}
 P(s)=\frac{1}{\beta_sN}\frac{\partial(\ln(\frac{C(V(s))}{N!}))}{\partial 
 {\bf s}}
 \label{e4} \end{equation}
 \begin{equation} F(s)=A(s)+{\bf S}P(s)=-\frac{1}{\beta_s}\ln(Z(s))+\frac{{\bf
 S}}{\beta_sN}
 \frac{\partial(\ln(\frac{C(V(s))}{N!}))}{\partial {\bf s}} \label{e5} \end{equation} 
Note that the forms of $P(s)$ and $F(s)$ are the same, except for the $s$
dependence, as those of the $N$ particle system \cite{Mayer}.   The boldfaced ${\bf S}$ and ${\bf s}$ in Eqs (\ref{e4})-(\ref{e5})  denote
 respectively the volumes in $s$  space available for $N$  and 
 one sites and the $A(s)$ in Eq (\ref{e5}) is the work function in this space 
 which 
  corresponds in its expression to  the analogous work function or the 
  Helmholtz free energy of
  the $N$ particle system \cite{Mayer,Reichl,Huang}. In order to be able to use
  and interpret the Quantity $P(s)$ (and $F(s)$), and especially to show that
  it is  related to the mentioned phase transition,  we have first to express
  $C(V(s))$ from Eqs (\ref{e2})-(\ref{e5}) in a form appropriate for discussing
  Internet websites. This form is found if we first ascertain the proper 
     numbering procedure suitable for the linked sites and which is taken care
     of by  
 the potential $V(s)$.  Thus, we  write this $V(s)$  as a sum of 
 terms  each of which 
 depends  on the ``distance''  $s_{ij}$  between any two sites $i$ and $j$ 
 which denotes,  as remarked,  the  
 connectivity  between them. Now, since each two sites may be connected to each
 other in three different ways as in the situations labelled $(1)$, $(2)$, and $(4)$ 
  in the 
 discussion after  Eq
 (\ref{e1})  we get the result that in a system of $N$ sites there are 
  $\frac{3N(N-1)}{2}$ 
 different pairs  so that the potential is
 \begin{equation} \label{e6} V(s)=\sum_{i\ne j}^{i=N}\sum_{j=1}^{j=N-1}u(s_{ij})+
 \sum_{i> j}^{i=N}\sum_{j=1}^{j=N-1}u(s_{ij})u(s_{ji})\end{equation} 
The first double sum takes account of the $N(N-1)$ pairs that are singly connected to
 each other and the second covers the $\frac{N(N-1)}{2}$ that are doubly
 connected and 
   $u(s_{ij})$ is the 
 potential of the pair of sites $i$ and $j$ as a function of the ``distance'' $s_{ij}$ 
  between them (see Eq (\ref{e1})). Note that the number of doubly connected pairs is
 half that of the singly connected ones since a single  connection
 between any two sites $i$ and $j$ may be realized in two different ways (see
 the situations $(1)$, and $(2)$  in the former discussion) compared 
 to the double connection 
between them which is realized in only one way.  
 We define, as done when discussing the cluster
 function theory \cite{Ursell,Mayer,Reichl}, the function $g_{ij}$ 
 \begin{equation} \label{e7} g_{ij}=g(s_{ij})=e^{-u(s_{ij})}-1  \end{equation} 
 We see   that $g_{ij}=0$ for a large ``distance'' $s_{ij}$ 
 (unrelated web sites) where $u(s_{ij})$ tends to zero (see Eq (\ref{e1})). Note  that when the sites $i$ and 
$j$ have no links to each other (unrelated sites) then they, naturally, also have 
no other common links so that $g_{ij}$ and $u(s_{ij})$ are both zero. 
Also, as one may realize  the
 probability to find two  sites that are entirely identical to each
 other is very small  so that $u(s_{ij})$ in this case becomes very large 
 (see Eq (\ref{e1}))  such
 that $g(s_{ij}) \approx -1$. By $g_{ij}$ we denote  the one-way connection
 from site $i$ to site $j$ when there is a link in $i$ to $j$,  and
 the double connection between them  is
 denoted by $g_{ij}g_{ji}$ where, obviously, we have (see Eqs (\ref{e1}) 
 and (\ref{e7})) 
$g_{ij} \ne g_{ji}$.   The ``configuration integral'' from Eq (\ref{e3}) 
becomes using the last equations 
 \begin{equation} \label{e8} C_N(V(s))=\int \int \ldots \int (1+\sum_{N \geq i \ne j
 \geq 1}g_{ij} +\sum_{N \geq i \ne j \geq 1}\sum_{N \geq \grave i \ne \grave j
 \geq 1} g_{ij}g_{\grave i \grave j}+\ldots) ds_1\ldots ds_N
  \end{equation} Note that due to the special character of the websites, as
  discussed after Eq (\ref{e1}), the  counting relation between  
  the general sites $i$ and $j$ is as written in the subscripts of the 
  summation signs of
  the last equation (compare with the analogous expression in the $N$ particle
  system (see  Eq (13.3) in p. 277 in \cite{Mayer})).  
  
  \protect \subsection{The plane diagrammatic structure of the linkings among
  the websites}
  
  \bigskip \bigskip 
  
Now, in order to be able to cope with the terms under the 
  parentheses signs in Eq (\ref{e8}) we use an extended version of the 
    one to one
correspondence  \cite{Mayer,Reichl} made between the terms of $C$ of the
$N$-particle ensemble 
   and certain planar diagrams. For example,  the  diagram shown in Figure 1 
  for $N=24$ corresponds to the term in Eq (\ref{e8}) that involves $g_{1,2}$, 
  $g_{3,4}$, $g_{5,6}$, $g_{9,10}$, $g_{11,12}$, $g_{11,19}$, $g_{12,19}$, 
 $g_{12,20}$, $g_{13,20}$,  $g_{15,16}$, $g_{15,23}$,  
  $g_{16,23}$, $g_{16,24}$, $g_{19,20}$ and  $g_{23,24}$  (we have inserted commas between 
the $i$ and $j$ components of $g_{ij}$). As seen from the diagram
  the sites (denoted in this paragraph by curly brackets) 
are assembled in clusters, so that the 
  sites \{7\}, \{8\}, \{14\}, and \{22\} each 
  constitutes a single cluster of 1 member. The sites \{1\}\{2\}, \{5\}\{6\}, 
  and \{17\}\{18\} are 
  in  clusters of two where the pairs \{1\}\{2\} and \{17\}\{18\} 
  are doubly connected. 
  The sites \{3\}\{4\}\{9\}\{10\} and \{15\}\{16\}\{23\}\{24\} are clusters 
  of 4 where the members of the first are singly connected to each other whereas
  those of the second are doubly connected.  
  The sites \{11\}\{12\}\{13\}\{19\}\{20\}\{21\} 
constitute a six-member  cluster.  As 
  seen  from Figure 1 the sites in a  cluster are connected in a 
  different manner to each other so that one may be 
 doubly  connected to all the other sites of the cluster in which case its
 connectivity is maximal whereas another may be singly 
  connected to only one site so that its connectivity is minimal.  
  For example, in the six member
  cluster of Figure 1 the site \{20\} is doubly connected to  the  
  sites \{11\},
  \{12\}, 
  and \{19\} and singly connected to \{13\} and \{21\},  
  whereas the sites \{13\} and \{21\} are each singly connected to only the 
  site 
   \{20\}. This difference between the 
    connected sites and the less connected ones is realized when one remove from
  the cluster one site and all its connecting lines.   Thus,  if the removed site 
  is the densely connected one the connectivity of the remained cluster is
  weakened considerably  whereas if the removed site  is slightly connected the
  effect on the connectivity of the remaining cluster is less influential. For
  example, if the site \{20\} and all its connecting lines in the six-member cluster
  of Figure 1 is removed the cluster is actually broken  into three different 
  smaller ones whereas if either the site \{13\} (or \{21\}) is removed together with 
  its single connecting line the connectivity of the remaining 5-member cluster
  is only slightly affected. \par 
  
  \begin{figure}[ht]
\centerline{
\epsfxsize=5.9in
\epsffile{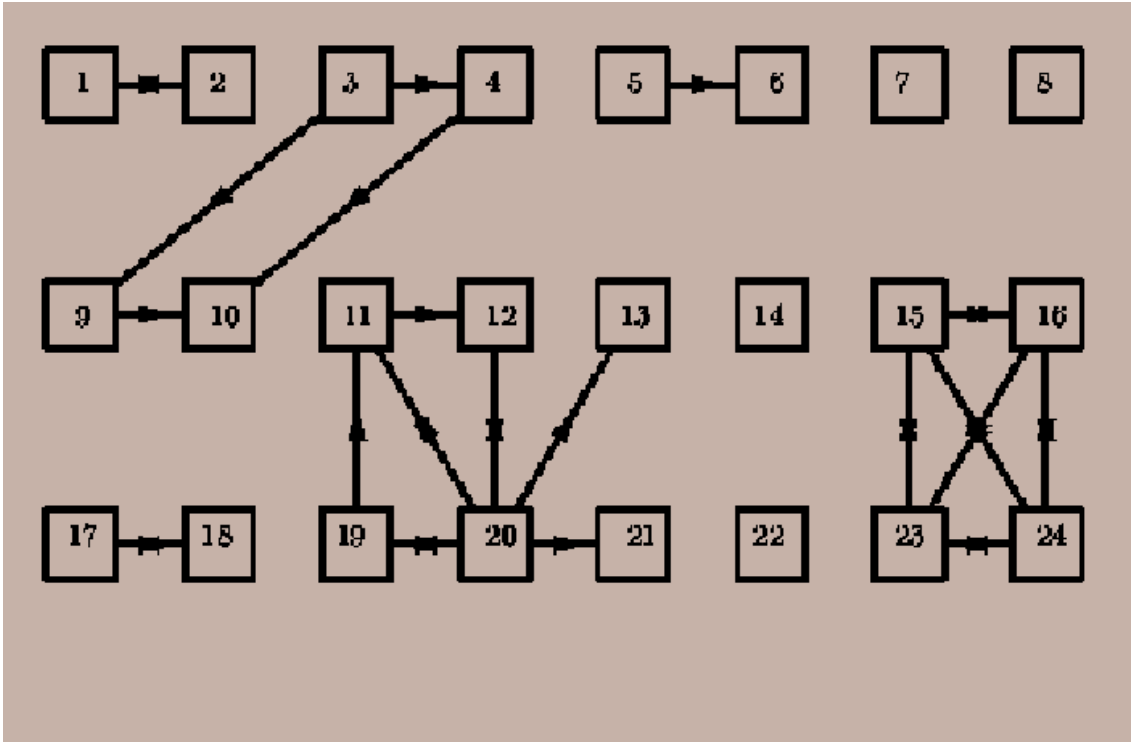}}
\caption[fegf4]{An example of the  planar diagram that maps the terms of the 
configuration integral  from Eq (\ref{e8}). Shown in the diagram   is  one
way, from a large number of  possible ones,  of connecting 24 sites into
clusters where the linking  between any two sites, if exists,  may be 
of the single connection type or the  double one. }
\end{figure}
  We note that Figure 1 is reminiscent of Fig 13.1 in page 278 in \cite{Mayer} which  
shows a similar diagram for 28 molecules. The principal difference between the two 
diagrams is that in Figure 13.1 in \cite{Mayer} the connection, if exists, between any 
two molecules is unique and not directed as in the plane diagram of Figure 1 here. 
That is, as remarked,  the linking between any two sites may be of 
three kinds and the lines that connect them reflect this  
diversity of connection.   

\protect \subsection{The ordering of the sums of the integrals of $C_N(V(s))$
into  clusters $c_l$}

\bigskip \bigskip

  We denote the number of times an $l$-site  cluster appears in a term by $m_l$
  so that the total number $N$ of all the sites may be written, as for the 
  analogous  
$N$ particle system   \cite{Mayer},  as 
  $N=\sum_{l=1}^{l=N}lm_l$.  Note that the integrals  over the sites in 
different clusters of one term  in Eq (\ref{e8}) are independent of 
each other.   Thus,  the integral of a term in Eq (\ref{e8}) is the product
  of the integrals over the sites in the same cluster where the 
  meaning of the integral 
over the $s$ variable is as discussed after Eq (\ref{e3}).      
    As for the $N$ particle system \cite{Mayer} we sum the integrals of 
    all the products that occur when a
  specified $l$-sites are in one cluster  and, because of the different counting
  of  Internet websites,  we divide this sum,  
  denoted by $c_l$, into two different parts as
  \begin{equation} \label{e9} 
  c_l=c_{l_m}+c_{l_d}=\frac{1}{l!{\bf S}}\int \int \ldots \int \sum 
  (\prod_{l \geq i\ne j \geq 1}g_{ij}+\prod_{l \geq i> j \geq 1}g_{ij}g_{ji})
  ds_1 \ldots ds_l, \end{equation}
 where $c_{l_m}$, $c_{l_d}$ denote the products over the mixed and doubly
  connected sites respectively.   By the term mixed we mean that the sites    
  in the relevant cluster may be either only  singly connected  or both singly 
  connected (to some sites)  and 
doubly  (to others, see the example of $c_3$ in Eq (\ref{e10})). Note the different counting relations between the general 
sites $i$ and $j$ for the mixed and doubly connected sites as expressed
respectively in the
subscripts of the two product signs of Eq (\ref{e9}). The counting relation of 
  $l \geq i> j \geq 1$ for the doubly connected sites, which is the same 
as in the analogous $N$
  particle system (see p. 277 in  \cite{Mayer}),  is because this kind of
  connection is unique and does not depend on direction.  
   The number of $g_{ij}$ in each term of the $c_{l_m}$ ranges from a
  minimum of $(l-1)$ to a maximum of $(l(l-1)-1)$ and the corresponding number in
  the $c_{l_d}$ part ranges from a minimum of $2(l-1)$ to a maximum of $l(l-1)$ where 
these numbers  are always even for $c_{l_d}$. 
   The product $l!{\bf S}$ 
  is a normalization factor where  ${\bf S}$ is, as remarked,  the total volume in $s$-space
 of all the $N$ sites.  
 For example, $c_3$ is 
\begin{eqnarray} && c_3=c_{3d}+c_{3m}=\frac{1}{3!{\bf S}}\int  \int \int 
     ds_1ds_2ds_3[(g_{31}g_{13}g_{21}g_{12}+g_{32}g_{23}g_{31}g_{13}+ \nonumber
     \\ &&+
g_{32}g_{23}g_{21}g_{12}+ 
    g_{32}g_{23}g_{31}g_{13}g_{21}g_{12}) +     (4g_{31}g_{21}+4g_{32}g_{31}
   +4g_{32}g_{21}+8g_{32}g_{31}g_{21}+ \nonumber \\ && + 
4g_{31}g_{13}g_{21}+  4g_{32}g_{23}g_{31}+ 4g_{21}g_{12}g_{32}+ 
      4g_{32}g_{23}g_{31}g_{21}
    +4g_{31}g_{13}g_{32}g_{21}+ \label{e10} \\ && +
    4g_{21}g_{12}g_{31}g_{32}+  
    2g_{31}g_{13}g_{21}g_{12}g_{32}
   + 2g_{32}g_{23}g_{31}g_{13}g_{21}+2g_{32}g_{23}g_{21}g_{12}g_{31})] \nonumber 
    \end{eqnarray} 
    The $c_{3d}$ is given by the first four terms and $c_{3m}$ is given by 
    the rest. 
    Note that since each mixed connection between any two sites may be,  as
    remarked, of three kinds there are 50 terms in $c_{3_m}$ whereas only 4   
     terms in $c_{3_d}$ because the double connection is unique.   We thus see
     that $c_l$ becomes very large for increasing values of $l$ as may be
     realized from Appendix A in which we use combinatorical analysis for
     calculating the number of terms in $c_l$ for general $l$. 
     
     \pagestyle{myheadings}
\markright{THE "CONFIGURATION INTEGRAL" $C_N(V(s))$, THE CLUSTER ....}

\bigskip \bigskip \bigskip 
     
     \protect \section{THE  "CONFIGURATION INTEGRAL"  $C_N(V(s))$, THE CLUSTER 
     INTEGRALS $c_l$ AND THE  IRREDUCIBLE  INTEGRALS $\zeta_k$}
     
     \protect \subsection{The expression of  $C_N(V(s))$ in terms of $c_l$} 
     
      \bigskip \bigskip
     
    From the discussion of the former section we realize that  
    the total contribution to the ``configuration integral'' from each specific    
 $l$-cluster is \cite{Mayer} 
\begin{equation} \label{e11} 
C(V(s))_{specific \ l-cluster}=\prod_l(l!{\bf S}c_l)^{m_l},  \end{equation} 
which is the expression 
obtained also for the $N$ particle system (see p. 280 in \cite{Mayer}) except 
that here 
we use the volume in $s$ space.    
The product
$(l!{\bf S})$ has been introduced to cancel the effect of the normalization constant
(see Eq (\ref{e9}) and the discussion following it)  and $m_l$ 
is, as noted,  the number
of times an $l$-cluster appears in a term.   
Now, we add all the similar contributions from the other $l$-site clusters 
which  is obtained by mutiplying the expression from the last equation by
$\frac{N!}{\prod_lm_l!l!}$ which is the number of ways to distribute $N$
different sites into clusters so that $m_1$ clusters have one site
each, $m_2$ clusters have two sites each ... and $m_l$ clusters have $l$ 
sites each. 
 The division by $\prod_lm_l!l!$ (and not by $\prod_lm_l!(l!)^{m_l}$ as for 
 the $N$ particle system 
(see page 281 in \cite{Mayer}))
 is because of the unique nature of the websites 
which are not only different from each other but also render their clusters,  
even those 
with the same number of sites, different. That is,  the division $\prod_lm_l!l!$
 takes into account  the 
permutations of the sites inside the clusters and also  
the permutations of the clusters that have identical number of sites. The 
overall 
contribution from all the $l$ site
clusters is therefore \begin{equation} \label{e12} C(V(s))_{all \ l-clusters}=
\prod_l\frac{(l!{\bf S}c_l)^{m_l}N!}{m_l!l!} =
\prod_l\frac{N!({\bf S}c_l)^{m_l}l!^{(m_l-1)}}{m_l!} \end{equation} 
The last step is to 
sum over all values of $m_l$ so that we obtain for the total value of the 
``configuration integral'' divided by $N!$ \begin{equation} \label{e13} 
\frac{C(V(s))}{N!}=
\underbrace{\sum_{m_l}\prod_l}_{\sum_{l=1}^{l=N} lm_l=N}\frac{({\bf s}Nc_l)^{m_l}l!^{(m_l-1)}}
{m_l!},  \end{equation} 
 where we have used the relation ${\bf s}=\frac{{\bf S}}{N}$ which is the volume in $s$
 space per site.  As seen from the last equation the indices $m_l$ and $l$
 over which the sum and the product are respectively run must satisfy the
 condition \cite{Mayer} $\sum_{l=1}^{l=N} lm_l=N$  (see the discussion before 
Eq (\ref{e9})).  
\protect \subsection{The ordering of the cluster integrals $c_l$  into
irreducible integrals $\zeta_k$}

\bigskip \bigskip

 As   realized from the last equations, there are many
 terms in $c_l$ which represent clusters that are composed of two parts that
 are connected by only one link, so that removing it splits the relevant integral
 into a product of two. Thus,  the cluster integrals $c_l$'s may be simplified 
 as 
  in the $N$ particle system  \cite{Mayer,Reichl}  by representing them as 
  sums of
 integrals that can not be further reduced into a product of integrals. That is,
 these irreducible integrals, denoted as $\zeta_k$, correspond to clusters 
 all the members of which are more than singly connected 
  as follows 
\begin{equation} \label{e14} 
 \zeta_k=\frac{1}{k!{\bf S}}\int \int \ldots \int \sum \prod_{k+1 \geq
  i\ne j 
  \geq 1}g_{ij}
ds_1 \ldots ds_{k+1}, \end{equation}  where, the product is over all the sites
each of which is   connected  to at least two other sites.  
Note that since, as remarked, the
connection between any two sites $i$ and $j$ may be of more than one kind 
 (see the discussion
after Eq (\ref{e1})) the counting relation is as written in the subscript of
the product sign (and not $k+1 \geq i> j  \geq 1$ which is appropriate for the
$N$ particle system (see p. 287 in \cite{Mayer})).   As for the $c_l$  
from Eq (\ref{e9}) we differentiate between 
$\zeta_{k_m}$ and $\zeta_{k_d}$ which denote mixedly and doubly connected sites respectively 
where by the term mixedly we mean that the sites in the relevant cluster may 
be, as for 
the $c_{l_m}$, either only singly connected or both singly connected (to some
sites) and doubly (to others).  
 From Eqs (\ref{e7}), (\ref{e9})  
    and (\ref{e14}) we see that the terms of $c_{l_m}$ and $\zeta_{k_m}$ may be positive or
    negative depending upon the evenness or oddness respectively of  the
    numbers of $g_{ij}$ in these terms. The terms of $c_{l_d}$ and $\zeta_{k_d}$, 
on the other hand, are always
    positive since the number of their $g_{ij}$ is always even. \par
The first two $\zeta_k$, for example,  are  
  \begin{eqnarray} && \zeta_{1}=\frac{1}{{\bf S}}\int \int
  (2g_{12}+g_{12}g_{21})ds_1ds_2 \nonumber  \\
  &&\zeta_{2}=\frac{1}{2{\bf S}}\int \int \int (6g_{32}g_{31}g_{21}+
  12g_{32}g_{23}g_{31}g_{21}+6g_{32}g_{23}g_{31}g_{13}g_{21}+ \label{e15} \\ 
 && +g_{32}g_{23}g_{31}g_{13}g_{21}g_{12}) ds_1ds_2 ds_3 
   \nonumber 
\end{eqnarray}
The coefficients of 2, 6,  and 12  signify the number of  similar terms that differ only by the direction  
 of the connecting lines  between the sites. Thus, the term $6g_{32}g_{31}g_{21}$ denotes 
the six possible different ways to construct a 3-cluster from three singly 
connected sites 
since each of these sites may be connected in two different ways to its neighbour. 
 Likewise, the term 
$12g_{32}g_{23}g_{31}g_{21}$ denotes the 12 different possible ways to 
construct a 
3-cluster from two doubly connected sites  which are singly connected to the 
third.  This 
is  because  there are 3 different ways to select the two doubly connected 
sites from  
three   and for each of these there are four different ways to singly connect 
these two 
sites to the third one. Now, we can write, analogously to the $N$ particle
system \cite{Mayer},  any cluster integral $c_l$ 
 as a sum of 
terms each of which depends upon the powers $n_k$ of the integrals $\zeta_k$ 
where  $n_k$ and $k$ are related by $\sum_{k=1}^{k=l-1}kn_k=l-1$. The $(-1)$ 
in the last relation is because one site is left over due to the definition of 
$\zeta_k$ (see the discussion before Eq (\ref{e14})).   
Thus,   using the  arguments  in the Appendix in  
  \cite{Mayer} (see p. 455-459 there) with respect to the $N$ particle system 
   we write the following dependence of $c_l$ upon $\zeta_k$  
\begin{equation} \label{e16} 
   c_l=\frac{1}{3l^2}\underbrace{\sum_{n_k}\prod_{k}}_{\sum_{k=1}^{k=l-1} kn_k=(l-1)}
  \frac{l^{n_k}\zeta_k^{n_k}k!^{(n_k-1)}}{n_k!},  \end{equation} 
  where we take  into
  account that the number of ways to distribute $l$ different objects into $n_k$
  clusters of $k$ objects each (with $\sum_{k=1}^{k=l-1} kn_k=(l-1)$) 
is $\frac{l!}{\prod_kk!n_k!}$.  
 The  expression (\ref{e16}) results also from considering  
  the  three possible connections between any two sites 
 that introduce different counting (see the discussion before Eq
  (\ref{e12}) and after (\ref{e15}) and note   
     the similarity between  Eqs (\ref{e13}) and 
  (\ref{e16})).   
  
  \protect \subsection{The general term of $C_N(V(s))$ as a function of $c_l$ and
  that of $c_l$ as a function of $\zeta_k$}
  
  \bigskip \bigskip
  
We follow in this subsection the same steps and the same expressions,   
used by Mayer   \cite{Mayer}   in
his demonstration of phase transition for the $N$ particle system,   except for
the $s$ variable and the  mentioned generalization necessary for 
discussing the Internet
websites.       
That is, we  demonstrate, using the following equations (\ref{e17})-(\ref{e23}),   
that the overall connectivity of the {\bf doubly}  
linked cluster may undergoes 
phase-transition. \par We begin by  using the Stirling
 approximation \cite{Abramov} for $\ln l!$, $\ln m_l!$,  $\ln k!$ and 
 $\ln n_k!$ so 
one may write the
  logarithms of the general terms  of   the sums of Eqs (\ref{e13}) and
  (\ref{e16}), denoted 
by $\ln \Gamma_c$ and $\ln \Gamma_{\zeta}$ respectively as   (compare with Eqs (13.13) and 
(14.2) in \cite{Mayer})   
  \begin{eqnarray}&& \ln \Gamma_c=\sum_{l=1}^{l=N}(m_l \ln ({\bf s}Nc_l)+(m_l-1)(l \ln l-l)
  -m_l \ln m_l+m_l) \nonumber \\ && \ln \Gamma_{\zeta}=\sum_{k=1}^{k=l-1}(n_k \ln 
  (\zeta_k)+(n_k-1) (k \ln k-k)+n_k \ln l-n_k \ln n_k+n_k)-  \label{e17} \\ && - 
  \ln 3-2\ln l 
  \nonumber \end{eqnarray} 
   Now, if all the $\zeta_k$'s are positive one
  may replace \cite{Mayer}, for large values of $l$, the logarithm of the 
   sums in Eqs (\ref{e13}) and  (\ref{e16}) by that of
  their largest terms.  These are   obtained  
    from Eqs (\ref{e17}) by  subtracting from the first of which  
    the constant  $(-\ln Z)$ multiplied by the condition 
     $\sum_{l=1}^{l=N} lm_l=N$ 
      and  from the second the constant $(-\ln \rho)$  multiplied  
    by
 the  condition   $\sum_{k=1}^{k=l-1} kn_k=(l-1)$.  These kinds of operation and 
the denotation of the constants as $(-\ln Z)$ and  $(-\ln \rho)$ are done, 
 as in  
\cite{Mayer},  for  a clear 
 representation   of the relevant calculations.   \par 
We, now,   differentiate 
 the first of the resulting expressions (related to $\ln \Gamma_c$)  
 with respect to $m_l$ 
  and the second (related to $\ln \Gamma_{\zeta}$)  with respect  to    $n_k$  
  and equate to 
zero so that    the values of $m_l$ and   $n_k$ that maximize 
  $\ln \Gamma_{b}$ and $\ln \Gamma_{\zeta}$ respectively are 
  \begin{equation} \label{e18} m_l={\bf s}Nc_ll^le^{-l}Z^l,  \ \ \ \\
  n_k=l\zeta_kk^ke^{-k}\rho^k \end{equation} 
   Note that,  due to the nature of $c_l$ and $\zeta_k$  
   (see Eqs (\ref{e9}) and (\ref{e14})),  
 both $Z$ and  $\rho$ have  dimension of inverse $s$.  
   Substituting for $n_k$ in the second of Eqs (\ref{e17}) and regarding the
   resulting expression, in the limit of very large $l$,  as representing   
   the
   logarithm of the sum in (\ref{e16}) one obtains, neglecting  
   the terms $(-\ln 3)$ and 
   $(-2\ln l)$
   compared to $l$ and using the condition for large $l$ $\sum kn_k=l$ 
(instead of $\sum kn_k=(l-1)$)
    \begin{equation} \label{e19} 
  \lim_{l\to \infty} \ln \Gamma_{\zeta}=\lim_{l \to \infty} \ln c_l=
  \lim_{l\to \infty}l(\sum_{k=1}^{k=l-1}
  \zeta_kk^ke^{-k}\rho^k-\ln \rho)=\lim_{l\to \infty}l\ln b_0, \end{equation} 
  where $\ln b_0=\sum_{k=1}^{k=l-1}\zeta_kk^ke^{-k}\rho^k-\ln \rho$,   
  which leads to
 \begin{equation} \label{e20} 
 b_0=\frac{\exp(\sum_{k=1}^{k=l-1} \zeta_kk^ke^{-k}\rho^k)}{\rho} \end{equation}
  Compare with the analogous expressions (14.6)-(14.7) in \cite{Mayer} for 
  the $N$ particle 
system. 
  From the last two equations one obtains, in analogy to the $N$ particle 
  system  
 \cite{Mayer} \begin{equation} \label{e21}  c_l=f(l,\zeta)b_0^l, \end{equation} 
     where $f(l,\zeta)$ is small compared
 to $l$ so that it satisfies $\lim_{l \to \infty}\frac{\ln f(l,\zeta)}{l}=0$.  
   Now, the condition $\sum_{l=1}^{l=N}lm_l=N$  may be written,  using the 
   first of Eq (\ref{e18}),  as
   \begin{equation} \label{e22} 
   \sum_{l=1}^{l=N}lm_l=\sum_{l=1}^{l=N}N{\bf s}l^{l+1}e^{-l}c_lZ^l=N  \end{equation}  
   Substituting  for $c_l$ from Eq (\ref{e21}) 
   into the last equation,    dividing by $N$, and using  for large $l$ the
   approximation   
   $l^{l+1}\approx l^l$  one obtains \begin{equation} 
  \label{e23} \sum_{l=1}^{l=N}{\bf s}f(l,\zeta)(le^{-1}b_0Z)^l=1 \end{equation} 
    The last equation, as shown in the next section, is central in demonstrating
     that enlarging the remarked  connectivity even slightly,  
   by adding the same 
links to them,  results in an enormous strengthening of
  the overall connectivity. 
  
  \pagestyle{myheadings}
\markright{THE PHASE TRANSITIONAL STRENGTHENING OF THE  ....}

\bigskip \bigskip \bigskip 

   \protect \section{THE PHASE TRANSITIONAL STRENGTHENING OF THE  CONNECTIVITY 
   AMONG  THE INTERNET  WEBSITES} 
   
   \bigskip \bigskip
   
   \protect \subsection{The results for a cluster of $l=N=10^5$ sites}
   
   \bigskip \bigskip
   
     We, now,   assume   that  the number of sites $l$ of the  
    $l$-cluster is large and that they    are doubly connected to 
   each other, in which
    case both $c_l$ and $\zeta_k$ are, as remarked,   positive due to 
   the even number of 
   their $g_{ij}$. Thus,  
        dividing both sides of Eq 
  (\ref{e23})  by ${\bf s}$ 
 and  taking the logarithm   of the term that corresponds
  to $l=N$ we obtain for  this term  (denoted  $\Gamma_N$)    
  \begin{equation} \label{e24} \ln \Gamma_N=\ln f(N,\zeta)+
  N\ln (Ne^{-1}b_0Z) \end{equation}  
   The expression $ Nb_0Ze^{-1}$ may be written as $e^{\epsilon}$ 
  so that if $\epsilon=0$  then   $Nb_0Ze^{-1}=1$, and when $\epsilon$ is 
  negative $Nb_0Ze^{-1}$  is smaller than unity and  the relevant term 
  $N\ln (Ne^{-1}b_0Z)$  in Eq (\ref{e24}) is negative.  In this case  since 
  it is the 
  dominant term    $\ln \Gamma_{N}$  on the left hand side of Eq (\ref{e24})  
  is also negative,  which implies that 
$\Gamma_N$  is 
  smaller than unity and  its contribution to the sum in Eq (\ref{e23})  
  is 
  small. Thus, for exemplifying the remarked giant increase of the connectivity    
 we assign  to $N$ the value of $N=10^{5}$ and take into account the
  former  result  of  $\lim_{l \to \infty}\frac{\ln f(l,\zeta)}{l}=0$, which 
  requires  $\ln f(N,\zeta)$ to be much smaller than
  $N$.         So,   if 
  $\epsilon$
  increases from zero by only the small amount of $10^{-4}$, due to adding 
a small Quantity of linked sites to the cluster,  the value of  
   $\ln \Gamma_{N}$ grows, as a result of this, by $10^1$  and 
   the contribution of $\Gamma_{N}$ to the  sum increases by
   $10^{10^1}$. That is,  adding even a  small number  of
 connecting  links to a large cluster of connected sites results in a  
   disproportional 
   strengthening of the  connectivity  of  this specific  cluster. \par 
The previous results may be realized from the two  
Subfigures  of Figure 2
which  show $\ln (\Gamma_N)$ and $\Gamma_N$ from Eq (\ref{e24}) as functions of
$\epsilon$ in the range   $-0.01 \le \epsilon \le 0.01$  and for  
  $N=10^{5}$.   
 The left hand side Subfigure  shows the change 
of $\ln (\Gamma_N)=N\epsilon$ from Eq
(\ref{e24}) in the  neighbourhood of $\epsilon=0$ where we have written
$Nb_0Ze^{-1}=e^{\epsilon}$ and ignore the term $\ln(f(N,\zeta))$  which vanishes
for large $N$. Thus, one may see that $\ln (\Gamma_N)$ is proportional to 
$\epsilon$ 
where the coefficient of proportion is $N$.  But the
behaviour of  $\Gamma_N=e^{N\epsilon}$ (compared to that of $\ln (\Gamma_N)$)  is not 
the same for
positive and negative $\epsilon$ as may be seen from the right hand side 
Subfigure  of Figure 2.
That is, although negative $\epsilon$ may produce  large negative values  of 
$\ln (\Gamma_N)$ as seen in the left hand side subfigure it yields  a rather negligible
change of $\Gamma_N$. But when $\epsilon$ departs slightly from zero towards positive
values the produced change in $\Gamma_N$ becomes so enormous 
  that even  the small change
of  
 $\delta \epsilon \approx
0.006$,  yields the 
giant change of $\Delta \Gamma_N \approx 10\cdot 10^{303}$. That is, the 
overall connectivity of the large cluster
has undergones, as remarked, a phase transition change. 

 \begin{figure}[ht]
\centerline{
\epsfxsize=5.9in
\epsffile{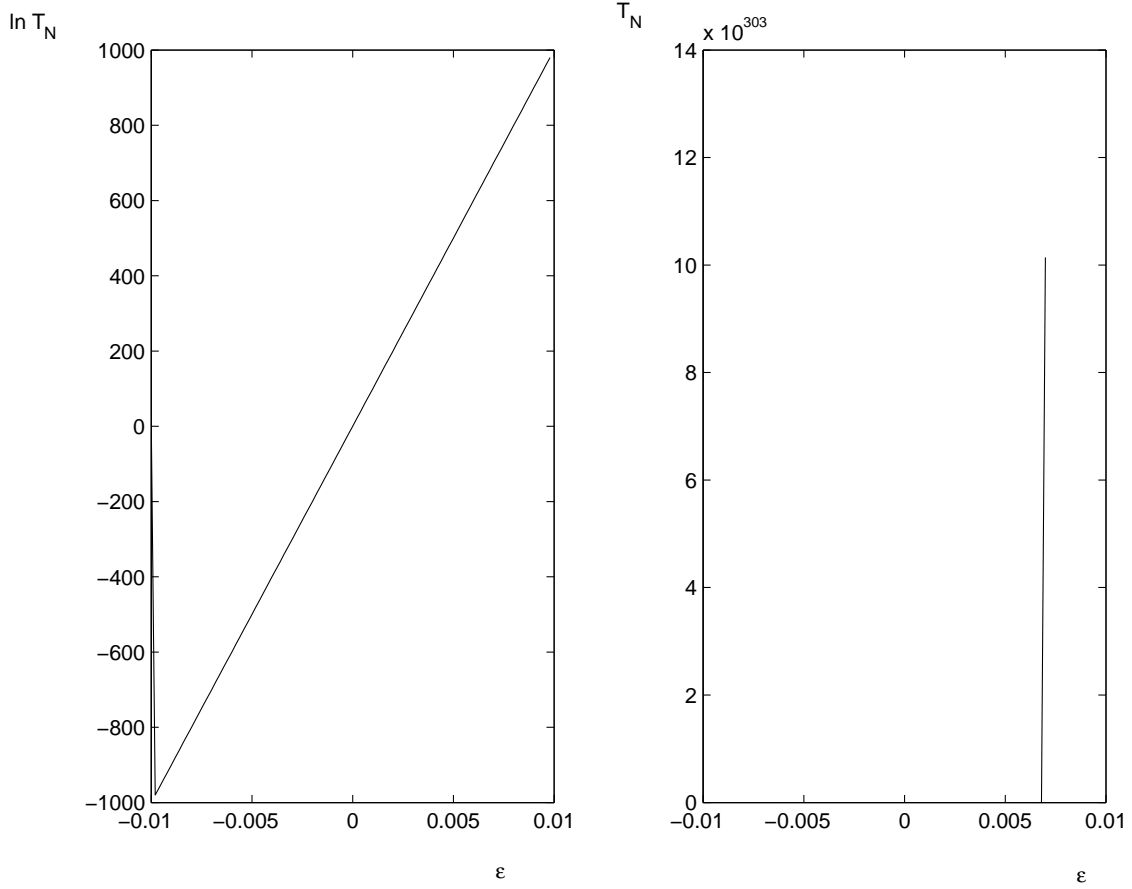}}
\caption[fegf4]{The left and right subfigures show respectively 
 the change of $\ln \Gamma_N$ and $\Gamma_N$ 
from Eq (\ref{e24}) in
the   neighbourhood of $\epsilon =0$.   Both graphs are plotted 
for  $N=100000$ and for the range of $\delta \epsilon=(-0.01, 0.01)$ where any
two neighbouring values of $\epsilon$ differ by $\frac{1}{5000}$. Note
the giant change of $\Delta \Gamma_N=10\cdot10^{303}$ for the slight increase of
$\delta \epsilon\approx 0.006 $. }
\end{figure}

 \protect \subsection{The inverse proportionality between the connectivity in a
 large $l$ cluster and the "volume" per site  in $s$ space}
 
 \bigskip \bigskip

       Now, remembering that Eq (\ref{e24}) was 
obtained after dividing both sides of Eq (\ref{e23})  by ${\bf s}$ we have  
to conclude,  
in order to
   retain Eq (\ref{e23}),  that ${\bf s}$ must decrease in this case to a correspondingly
   very small value 
    where, as noted,  ${\bf s}$ is the "volume" in $s$ space per site.  
For a better understanding of the meaning of a small ${\bf s}$ we return to Eq
(\ref{e4}) for the "pressure"  $P(s)$ and evaluate it in the limit of large  
$l$. In such case one may replace, as remarked,   $\ln (\frac{C(V(s))}{N!})$,  
where  $\frac{C(V(s))}{N!}$ is given by Eq (\ref{e13}), 
by $\ln (\Gamma_c)$ from the first of Eqs (\ref{e17}). Thus, using
$m_l={\bf s}Nc_ll^le^{-l}Z^l \approx {\bf s}Nc_ll^{(l+1)}e^{-l}Z^l$ and $\sum lm_l=N$ (see
the first of Eqs (\ref{e18})  and Eq (\ref{e13})) we may evaluate  
the "pressure" 
$P(s)$ from Eq (\ref{e4}) as follows  \begin{eqnarray} && 
P(s)=\frac{1}{\beta_sN}\frac{\partial(\ln(\frac{C(V(s))}{N!}))}{\partial {\bf s}} =
\frac{1}{\beta_sN}\frac{\partial(\ln(\Gamma_c))}{\partial {\bf s}}=  \label{e25} \\ 
&& = \frac{1}{\beta_sN}
\frac{\partial (\ln(\sum_{l=1}^{l=N}(m_l(\ln ({\bf s}Nc_l)+(l \ln l-l)-\ln m_l+1)-
l(\ln l-1))))}{\partial {\bf s}}
= \nonumber \\ && = \frac{1}{\beta_sN}
\frac{\partial (\ln (\sum_{l=1}^{l=N}({\bf s}Nc_ll^le^{-l}Z^l(1-
\ln(l^le^{-l}Z^l)+l(\ln l-1))-l(\ln l-1))))}{\partial {\bf s}}= \nonumber \\ 
 && = \frac{1}{\beta_sN}
\frac{\partial (\ln (\sum_{l=1}^{l=N}({\bf s}Nc_ll^le^{-1}Z^l-l(\ln l-1))-
N \ln Z))}{\partial {\bf s}}=\frac{1}{\beta_sN}\frac{N}{{\bf s}}=
\frac{1}{\beta_s{\bf s}} \nonumber 
\end{eqnarray}
Substituting  the last result in the expression for the "free energy" from Eq 
(\ref{e5}) we obtain 
$ F(s)=A(s)+{\bf S}P(s)=-\frac{1}{\beta_s}\ln(Z(s))+\frac{{\bf S}}{\beta_s{\bf
s}}$. 
 Note that although $c_l$ contains the  total "volume" ${\bf S}$ in its denominator 
 (see Eq (\ref{e9})) 
 it certainly does not depend on it since  the integrals of $c_l$ always
 lead to a factor ${\bf S}$ that cancels that  in the denominator.   
Thus, from the last equations we see that 
the "pressure" $P(s)$ (and the "free energy" $F(s)$) becomes very large 
for very 
small ${\bf s}$.  
The meaning of small ${\bf s}$ may now  be understood by 
  recalling  that  the sites of the relevant linked cluster 
 are related to each other by the variable $s$ that denotes the 
``distance'' between them in the sense of  how much they are similar to  
each other (see the 
discussion before Eq (\ref{e1})). That is,  small ${\bf s}$ 
 for the 
large cluster of doubly linked sites means  that the "distances"  that signify the 
differences between them become also very small and they all turn out to be similar 
to each other. The vanishing "distances" between the sites in this case results
in  high values for the  "pressure" and the "free energy"  (see the last
equations) of these "jammed"
sites.  This occurs, as remarked, when    
 adding  a small amount of links  to the larger terms of
   the sum in Eq (\ref{e23})  which results in the outcome that the 
   overall connectivity 
among the component sites becomes maximal in the sense that they have the 
same links and 
so they are similar to each other.  \par 
  Note that  the former discussion depends on the assumption that all the 
$\zeta_k$ are positive  whereas we know that  
the signs of $c_l$ and $\zeta_k$ alternate due to the evennes and  
oddness of the 
number of $g_{ij}$ (see the discussion after Eq (\ref{e14})). But if we confine our 
discussion to only the doubly connected sites then all the $c_l$ and $\zeta_k$ are, 
as remarked, positive (in addition to their being mutually connected as required) 
so  all the  former discussion    
is obviously valid. That is, the doubly connected clusters may  
show this kind of phase 
transition-like of the  
 connectivity by only  adding a very small amount of connecting links. 
Note that the mathematical expressions of the doubly connected $N$ site system  
are  the same as the $N$ particle 
system discussed in \cite{Mayer}
  except that in the websites system  all the $c_{l_d}$ and $\zeta_{k_d}$ are 
  positive whereas they alternate in sign for the particle system.  Also,  
 this kind of avalanche (or condensation as termed in \cite{Mayer}) have been 
 shown in 
\cite{Mayer} for the $N$ particle system  provided  one assumes 
that all the $\zeta_k$ 
are positive (as here).  \par  
 We note that the process just described of a large increase in the fraction
 of the connectivity  in linked  large-size clusters is frequently encountered 
   in ensembles of shared computers connected to each other and to the Internet.
   In this case an additional web site in the form of a file that has been
   programmed intentionally for the purpose of adding and connecting it to the
   other files of the shared ensemble  enhances
   the connectivity of these members to a very high degree. For example, suppose
   that all these members use some utility program and that not all of them have
   the same version of it but  use several ones.  Thus, 
   adding an updation file of this program to them  turns all the
   different versions into one that is common to all thereby increasing very
   much the total connectivity of them. 

    \par 
     
  \bigskip \bigskip \noindent
  
\protect \section{CONCLUDING  REMARKS \label{sec10}}

We have discussed the Internet statistics using a generalized version of the 
Ursell-Mayer cluster formalism \cite{Ursell,Mayer,Reichl}. We   
 discuss the relevant elements  of the Internet, 
which are  the 
websites, by introducing a new variable that takes account of the unique 
nature of 
them. The special character of the Internet, unlike any other common  
fractal web, 
is especially 
demonstrated through the links in its websites which may be of  
several kinds as discussed 
after Eq (\ref{e1}). Taking into account this unique character  and 
introducing the relevant expressions into the Ursell-Mayer framework we have 
shown that 
one may obtain a phase-transition phenomena. This occurs  when a large 
cluster of doubly linked sites are added an extra small amount of connecting  
links in 
which case the contribution of this cluster to the overall connectivity of the entire 
ensemble becomes enormous. These phase transition changes were also seen to
characterize the "pressure" $P(s)$ and the "free energy" $F(s)$. 
    
 \protect \section{ACKNOWLEDGEMENT \label{sec11}} 
I wish to thank L. P. Horwitz for discussion on this subject and 
for reading the manuscript. 

\newpage 

\pagestyle{myheadings}
\markright{THE COMBINATORICS OF THE $l$-SITE CLUSTER} 

\vspace{2cm}

\begin{appendix}

\bigskip \bigskip

\protect \section{APPENDIX:  THE COMBINATORICS OF THE $l$-SITE CLUSTER \label{sec12}}

     We  calculate in this Appendix   the number of terms contributed  to the ``configuration 
    integral'' by any {\it specific $c_l$}.  
 This contribution is
  valid only when the  sites in the same cluster are linked  
  to each other and  
   is obtained by exploiting the special character of the $c_l$ from Eq 
  (\ref{e9})  in order to find all the possible ways by which
  it contributes to the configuration integral. We take into account that the
  number of $g_{ij}$ in the  terms  of $c_l$   ranges, as noted,  from $(l-1)$ to
  $l(l-1)$ and that  each $g_{ij}$  denotes, as remarked,   a link to site $j$ in site
  $i$. Thus,  the contribution of each $c_l$ may be obtained by calculating the number
  of possible ways by which each of the Quantities of links $(l-1), l, \ldots l(l-1)$, 
  may  link $l$ sites among them  
so as to construct an $l$-site cluster.  
We  begin from the least linked $l$-cluster 
which  is constructed by using only $(l-1)$ different links.  The number of ways
to construct such cluster   is (we denote this number $W_l(l-1)$) 
$$  W_l(l-1)=\left( \begin{array}{c} l(l-1) \\ (l-1) 
\end{array} \right)=\frac{(l(l-1))!}{(l-1)!(l-1)^2)!} \eqno (A.1)$$ 
$\left( \begin{array}{c} l(l-1) \\ (l-1) \end{array} \right)$  in the middle 
expression is  the number of combinations required to construct this least
$l$-site cluster where there are no two links in any of these combinations that
refer to each other such as $g_{ij}$ and $g_{ji}$.   
 Now, since, as remarked, the number of $g_{ij}$ in the terms of Eq (\ref{e9}) 
 ranges from $(l-1)$ to $l(l-1)$ we have, in order to find all the possible
  different contributions of each  $c_l$, to calculate also the $W_l(l), 
   W_l(l+1) \ldots W_l(l(l-1)) $. 
  Each of these is found by first constructing the minimally connected 
 $l$-cluster from $(l-1)$ sites as in Eq (A.1) and   since each
    site in the $l$-cluster may be  linked  to all the other $(l-1)$ sites,   except 
to itself, 
    the next step is to  link each of the sites    
    to all the others.   That is, the total number of ways to compose an $l$
cluster using a number of links  that ranges from $(l-1)$ to   $l(l-1)$
is  
 $W_l(l-1)+W_l(l)+\ldots +W_l(l+k)+\ldots W_l(l(l-1))$.  
   $W_l(l-1)$ in the former sum is given by 
Eq (A.1) and each one of the other $W$'s is obtained by noting that 
 after composing the minimally
linked $l$-cluster using the $(l-1)$ links, from the possible $l(l-1)$, 
 one remains with 
$l(l-1)-(l-1)=(l-1)^2$  links. These links, in contrast to the former
least linked $l$ cluster, may be formed so that it is allowed to count also
mutually linked sites such as $g_{ij}$ and $g_{ji}$. That is, for calculating
the number of ways to link these $(l-1)^2$ additional links we may   permute them
instead of the former combinations used for the initial $(l-1)$ links. Thus, as
remarked, for calculating $W_l(l), \ldots W_l(l+k), \ldots  W_l(l(l-1))$ 
we first construct the least linked $l$ cluster,  using $(l-1)$ links as in Eq
(A.1), and then we link the remaining links  which range from 1 for $W_l(l)$ to 
$(l-1)^2$ for $W_l(l(l-1))$.  For example, the
number of possible ways to construct an $l$ site cluster using $(l+k)$ links,
where $0 \le k \le l(l-2)$  is given by the recursive relation 
$$ 
W_l(l+k)=W_l(l+k-1)\cdot (l(l-1)-(l+k-1))=W_l(l+k-1)\cdot ((l-1)^2-k) \eqno 
(A.2) $$ 
The factor  $(l(l-1)-(l+k-1))$ in the middle expression is the number of ways 
to link the remaining links, from the initial possible $l(l-1)$,  
after linking $(l+k-1)$ links. When $k=0$ we obtain  from Eq 
(A.2) 
 $W_l(l)=W_l(l-1)\cdot (l-1)^2$.  From the last equation one obtains    the
number of ways for using $l$ links, from a total of $l(l-1)$, to construct an $l$
cluster where the first $(l-1)$ links are used to initially compose the least 
linked cluster and the additional link may be anyone from the remaining 
  $(l-1)^2$. When $k=l(l-2)$ one obtains from Eq (A.2) 
  $W_l(l(l-1))=W_l(l(l-1)-1)$.     
 Now, we can calculate the sum $W_l(l-1)+W_l(l)+\ldots+ W_l(l+k)+\ldots+  
 W_l(l(l-1))$ in order to find the number of ways by which one may build an
 $l$-site cluster. That is, we  write, using the former equations 
 $$  W_l(l-1)+W_l(l)+\ldots+ W_l(l+k)+\ldots+  
 W_l(l(l-1))= $$ 
 $$ = W_l(l-1)[1+(l-1)^2[1+((l-1)^2-1)+
 ((l-1)^2-1)((l-1)^2-2)+ $$ 
  $$ +\ldots + 
 \prod_{m=1}^{m=k}((l-1)^2-m)+\ldots+
\prod_{m=1}^{m=l(l-1)}((l-1)^2-m)]]=  \eqno (A.3) $$  
$$ = 
 W_l(l-1)[1+(l-1)^2[1+\sum_{k=1}^{k=l(l-1)}\prod_{m=1}^{m=k}((l-1)^2-m)]]  
  $$ 
 The last result is the number of terms of $c_l$ and it is very large for large
  $l$. 
  
 \end{appendix}
 
 \pagestyle{myheadings}
\markright{REFERENCES}

\bigskip \bibliographystyle{plain}

\end{document}